\documentclass[sigplan, nonacm, 10pt]{acmart}

\usepackage{algorithm}
\usepackage{algpseudocode}
\usepackage{xspace}
\usepackage{subcaption}

\acmConference[PPoPP'20]{ACM SIGPLAN Annual Symposium on Principles of Parallel Programming}{February 22--26, 2020}{San Diego, CA, USA}
\acmYear{2020}
\acmISBN{} 
\acmDOI{} 
\startPage{1}

\newcommand{\distSpeedup}{604$\times$\xspace}
\newcommand{\distSpeedupOverGreedy}{1400$\times$\xspace}
\newcommand{\distBaseline}{39314\xspace}
\newcommand{\distStrongScaling}{90\%\xspace}
\newcommand{\distSmallStrongScaling}{79\%\xspace}
\newcommand{\sharedSpeedupOverGreedy}{60.2$\times$\xspace}
\newcommand{\sharedSpeedup}{21.5$\times$\xspace}
\newcommand{\wittwerBaseline}{89486\xspace}
\begin{document}

%
\title[Parallel and Scalable Precise Clustering]{Parallel and Scalable Precise Clustering for Homologous Protein Discovery}

%
\author{Stuart Byma}
\email{stuart.byma@epfl.ch}
\affiliation{%
  \institution{EPFL}
}
\author{Akash Dhasade}
\email{akashdhasade@gmail.com}
\affiliation{
  \institution{IIT}
}
\author{Adrian Altenhoff}
\email{adriaal@inf.ethz.ch}
\affiliation{%
  \institution{ETH Z\"urich}
}
\author{Christophe Dessimoz}
\email{christophe.dessimoz@unil.ch}
\affiliation{%
  \institution{University of Lausanne}
}
\author{James R. Larus}
\email{james.larus@epfl.ch}
\affiliation{%
  \institution{EPFL}
}

%
\renewcommand{\shortauthors}{Byma, Dhasade, Altenhoff, Dessimoz, Larus}

\newcommand{\sbx}[1]{\noindent{{\bf \fbox{SB:} {\textcolor{red}{\it#1}}}}}
\newcommand{\jl}[1]{\noindent{{\bf \fbox{JL:} {\textcolor{red}{\it#1}}}}}

%
\begin{abstract}
  This paper presents a new, parallel implementation of clustering and demonstrates its utility in greatly speeding up the process of identifying homologous proteins.
  Clustering is a technique to reduce the number of comparison needed to find similar pairs in a set of $n$ elements such as protein sequences.
  \textit{Precise clustering} ensures that each pair of similar elements appears together in at least one cluster, so that similarities can be identified by all-to-all comparison in each cluster rather than on the full set.
  This paper introduces ClusterMerge, a new algorithm for precise clustering that uses transitive relationships among the elements to enable parallel and scalable implementations of this approach. 

  We apply ClusterMerge to the important problem of finding similar amino acid sequences in a collection of proteins.
  ClusterMerge identifies 99.8\% of similar pairs found by a full $O(n^2)$ comparison, with only half as many operations.
  More importantly, ClusterMerge is highly amenable to parallel and distributed computation.
  Our implementation achieves a speedup of \distSpeedup on 768 cores (\distSpeedupOverGreedy faster than a comparable single-threaded clustering implementation), a strong scaling efficiency of \distStrongScaling, and a weak scaling efficiency of nearly 100\%.
\end{abstract}

%
%


\settopmatter{printfolios=true}

%
\maketitle


\section{Introduction}\label{sec:intro}
The ongoing revolution in genome sequencing is generating large and growing datasets whose value is exposed through extensive computation.
The cost of this analysis is an impediment to analyzing these databases when the time for processing grows rapidly as a dataset becomes larger, more inclusive, and more valuable.  
Asymptotically efficient algorithms are desirable, but sometimes a tradeoff between speed and precision requires the use of expensive algorithms.
In this situation, the time to perform an analysis can be reduced by running on a parallel computer or cluster of computers.
This paper describes a new approach to applying parallel computing to protein clustering, an important technique in the field of \textit{proteomes}, the analysis of the protein sequences contained in an organism's genome.

Similarities among protein sequences are used as proxies to infer common ancestry among genes. 
Similar genes are referred to as \textit{homologs}, and their detection allows the transference of knowledge from well-studied genes to newly sequenced ones.
Homologs, despite having accumulated substantial differences during evolution, often continue to perform the same biological function. 
In fact, most of today's molecular-level biological knowledge comes from the study of a handful of model organisms, which is then extrapolated to other life forms, primarily through homology detection. 
Several sequence homology techniques are among the 100 most-cited scientific papers of all time~\cite{van_noorden_top_2014}. 

Current approaches to find similar (homologous) proteins are computationally expensive. 
The baseline is to perform an exhaustive, all-against-all ($O(n^2)$) comparison of each sequence against all others using the Smith-Waterman (S-W) or another, similarly expensive ($O(n^2)$) string matching algorithm.
This naive approach finds all similar pairs, but it scales poorly as the number of proteins grows.
Several databases of similar proteins produced by this approach exist, including OMA~\cite{altenhoff_oma_2018} and OrthoDB~\cite{waterhouse_orthodb:_2013}.
Analyzing their contents is costly.
OMA for example has consumed over 10 \textit{million} CPU hours, but includes proteins from only 2000 genomes. 

The large amount of data produced by many laboratories requires new methods for homology detection.
In a report published in 2014, the \textit{Quest for Orthologs} consortium, a collaboration of the main cross-species homology databases, reported: ``\textit{[C]omputing orthologs between all complete proteomes has recently gone from typically a matter of CPU weeks to hundreds of CPU years, and new, faster algorithms and methods are called for}''~\cite{sonnhammer_big_2014}. 
Ideally, a new algorithm with asymptotically better performance would find the same similarities as the ground truth, all-against-all comparison.
Unfortunately, fast (sub $O(n^2)$) algorithms --- based on k-mer counting, sequence identity, or MinHash --- identify significantly fewer homologs and hence are not practical for this application.
In the absence of a better algorithm, a scalable parallel implementation of an $O(n^2)$ solution would help  keep pace with the production of sequence data.

Our approach extends the idea of clustering~\cite{wittwer_speeding_2014} into \textit{precise clustering}, which ensures that each pair of similar proteins appears together in at least one cluster\footnote{Since a protein sequence can be in more than one cluster, clustering is not partitioning.}.
Similar pairs are then easily identified in the resulting clusters.
Traditional clustering techniques such as k-means, hierarchical clustering, density/spatial clustering, etc. are difficult to apply because they partition, require a similarity matrix,
or generally do not achieve sufficient selectivity.
Our technique uses the transitivity of similarity to construct clusters and to avoid unnecessary comparisons. 
The key idea is that some \emph{similar} sequence pairs will have the stronger property of \emph{transitively similarity}.
Formally, if $A$ is transitively similar to $B$, and $B$ is similar to $C$, then $C$ will be similar to $A$. 
This not only finds similarity between sequences $A$ and $B$, but also between $A$ and all sequences similar to $B$.
Transitivity avoids a large number of comparisons and reduces the computational cost.

We exploit transitivity by building clusters of sequences centered on a \emph{representative} sequence to which all cluster members are similar.
Any sequence \emph{transitively similar} to a cluster representative is added to its cluster.
It need not be compared against the other cluster members, as transitivity implies its similarity with them.
Sequences that are only \emph{similar} to the representative are also added to the cluster, but they must also be made representatives of their own, new cluster to ensure they are available for subsequent comparisons.
In this way, all similar pairs end up together in at least one cluster.
Previous work showed that this approach performs well for protein clustering~\cite{wittwer_speeding_2014}, but the greedy implementation in that paper was slow and not scalable.

In this paper, we generalize the problem of clustering with a transitive relation, introduce a parallel and distributed algorithm, and apply our approach to clustering protein sequences.
Our new algorithm for precise clustering is called \textit{ClusterMerge}.
The key insight enabling parallelism is that two clusters can be \textit{merged} if their representatives are transitively similar since each cluster's members are similar to the other cluster's representative (and members).
Members of both clusters can be merged into a single cluster with one representative.
If the representatives are similar (but not transitively similar), the clusters exchange elements that are similar to the other cluster's representatives.
The result of merging is either one cluster or two \emph{unmergeable} clusters (since their representatives are not transitively similar). 
Merging clusters reframes clustering as a process that starts with single-element clusters containing each element in the dataset and merges them bottom-up until a set of unmergeable clusters remains.

ClusterMerge exposes a large amount of parallelism in its tree-like computation.
However, the computation is highly irregular because of the wide span in the length of proteins (hundreds to tens of thousands of amino acids), the $O(n^2)$ string comparison that exaggerates this disparity, and differences in the size of clusters, all of which requires dynamic load balancing to achieve good performance.
We present efficient parallel and distributed implementations using this cluster merge approach.
Our single-node, shared-memory design scales nearly linearly and achieves a speedup of 21$\times$ on a 24 core machine.
Our distributed design achieves a speedup of \distSpeedup
while maintaining a strong scaling efficiency of \distSmallStrongScaling on a distributed cluster of 768 cores (\distStrongScaling on larger datasets), running \distSpeedupOverGreedy faster than the incremental greedy clustering of Wittwer et al.~\cite{wittwer_speeding_2014}.
Our distributed implementation exhibits a weak scaling efficiency of nearly 100\% on 768 cores. 
ClusterMerge and our implementations for protein sequence clustering are open-sourced~\cite{noauthor_clustermerge_nodate}

This paper makes the following contributions:

\begin{itemize}
  \item A formalization of precise clustering using similarity and transitivity. 
  \item An algorithm, ClusterMerge, that reformulates the clustering process in a parallelism-friendly form.
  \item An application of ClusterMerge to the problem of clustering protein sequences that maintains near-perfect accuracy while achieving high parallel efficiency.

\end{itemize}

The rest of this paper is organized as follows:
\S\ref{sec:related} reviews related work in clustering and sequence clustering.
\S\ref{sec:algo} formalizes precise clustering and presents the ClusterMerge algorithm.
\S\ref{sec:protclust} shows how to apply ClusterMerge to precise protein sequence clustering.
\S\ref{sec:impl} discusses our shared memory and distributed implementations of ClusterMerge.
\S\ref{sec:eval} evaluates the algorithm, systems, and their performance in this application. 
\S\ref{sec:future} discusses future work and \S\ref{sec:conclusion} concludes.
\section{Related Work}\label{sec:related}
Clustering in general has been the subject of considerable research.
Andreopoulos et al.\ survey uses of the techniques in bioinformatics~\cite{andreopoulos_roadmap_2009}.
Widely known techniques are difficult to apply to protein clustering, however.

Partitioning algorithms require an equivalence relation between elements, which is stronger than the not-necessarily transitive similarity relationship in protein clustering.
k-means clustering requires a target number of clusters, which is unknown in advance for proteins, and partitions the set.
Hierarchical methods partition elements into a tree and preserve hierarchy among elements, but generally require a similarity matrix to exist, which is not the case for our problem, and are expensive ($O(n^3)$).
Of particular note is \textit{agglomerative} hierarchical clustering, which also uses bottom-up merge, e.g., ROCK~\cite{guha_rock:_2000}.
Density-based clustering uses a local density criterion to locate subspaces in which elements are dense; however, they can miss elements in sparse regions and generally cannot guarantee a precise clustering.
Density-based techniques have received attention from the parallel computing community, with the DBSCAN~\cite{sander_density-based_1998} and OPTICS~\cite{ankerst_optics:_1999} algorithms being parallelized by Patwary et al.~\cite{patwary_new_2012, patwary_scalable_2013}

An additional complication of these methods is that they rely on distance metrics in normed spaces, e.g., Euclidean distance, which are usually cheap to compute.
Edit distance however, is not a norm and is expensive to compute.
Although pure edit distance (i.e., Levenshtein distance) can be embedded in a normed space~\cite{ostrovsky_low_2007}, it is not clear if the gapped \textit{alignment} necessary for protein similarity can be as well.

Clustering of biological sequences is the subject of considerable research.
Many of these clustering algorithms employ iterative greedy approaches that construct clusters around representative sequences, a sequence at a time.
If the sequence is similar to a cluster representative, it is placed in that cluster.
If the sequence is not similar to any existing cluster representative, a new cluster is created with the input sequence as its representative.
Some approaches use k-mer counting to approximate similarity (CD-HIT~\cite{li_cd-hit:_2006}, kClust~\cite{hauser_kclust:_2013}, Mash~\cite{ondov_mash:_2016}), 
while others use sequence identity, i.e., the number of exact matching characters (UCLUST~\cite{edgar_search_2010}).
Of note is Linclust~\cite{steinegger_clustering_2018}, an approach that operates in linear time
by selecting \textit{m} k-mers from each sequence and grouping sequences that share a k-mer.
The longest sequence in a group is designated its \textit{center} and other sequences are compared against it, avoiding a great deal of computation. 

Unfortunately, sequence identity and k-mers are unsuitable for finding many homologs.
Protein alignment substitution matrices are heterogeneous (e.g., BLOSUM62~\cite{henikoff_amino_1992}) since distinct amino acids may be closely related.
Hence, protein sequences that appear different --- with low sequence identity and therefore few or no shared k-mers --- can often have high alignment scores. 
These similar pairs will be missed by k-mer-based clustering techniques.
For example, the fraction of similar sequence pairs found by kClust, UCLUST, MMSeqs2 linclust, and MMSeqs2 are 10.4\%, 13.5\%, 0.5\%, and 36.4\%, respectively.

Wittwer et al.~\cite{wittwer_speeding_2014} use an iterative greedy approach to cluster protein sequences, using transitivity in the data to avoid comparing each sequence with all others while recovering $\sim$99.9\% of similar pairs.
Our work uses a similar transitivity function.
However, the previous iterative greedy approach is slow and difficult to parallelize because each added sequence depends on the clusters from the previous sequences and requires fine-grained synchronization.

\section{Precise Clustering}\label{sec:algo}

$S$ is a set of elements.
Elements in $S$ can be compared using a similarity function $f(i, j)$ that returns a measure of the similarity between elements $i$ and $j$.
We wish to find all element pairs $(i, j)$ in $S$ such that $f(i, j) > T$, where $T$ is a threshold parameter. 
We call these pairs \textit{significant pairs}.

While significant pairs can be found by pairwise of comparison of all elements in $S$, this requires $O(n^2)$ comparisons.
To avoid examining \textit{all} possible pairs, we cluster elements in $S$ such that for all element pairs $(i, j)$ in $S$ where $f(i, j) > T$, both $i$ and $j$ are members of at least one cluster. 
We call this a \textit{precise clustering}.
A cluster $C$ is a subset of $S$ defined as follows:
\begin{equation}\label{eq:cluster}
\forall e \in C, f(e, r_C) > T 
\end{equation}
where $r_C$ is the unique \textit{representative} element of the cluster.

The similarity function $f$ is not an equivalence function --- elements in a cluster are similar to its representative, but not necessarily to each other (although that may be likely).
In addition, a representative is not required to be similar to other elements similar to its cluster members.
To identify all significant pairs in $S$, each element would need its own cluster, and the problem devolves to all-against-all comparison.

Therefore, to avoid this, we exploit a stronger property of \textit{transitive similarity}.
Formally, for elements $i, j$ and $k$, if $i$ is transitively similar to $j$, and $i$ is similar to $k$, then $j$ is similar to $k$.
Therefore, if $i$ is the representative of a cluster, then we can infer similarity between $j$ and every other element in the cluster.
We require an additional \textit{transitivity function} $R(i, j)$ defined:
\begin{equation*}
\forall (i, j, k) \in S, R(i, j) \implies f(i, k) > T \land f(j, k) > T
\end{equation*}
$R(i, j)$ tells us that elements $i$ and $j$ are \textit{transitively similar}.
Element $j$ can be clustered with element $i$ and its similarity with other cluster members ($k$) inferred.
This reduces the number of comparisons needed to form a precise clustering.

\subsection{Merging Clusters}

Our key to exposing parallelism lies in recognizing that clusters with transitively similar representatives can be \textit{merged}.
This allows us to reframe clustering as a series of \textit{cluster merges}.
Two clusters can be merged as follows.
First, the representatives are compared using the similarity function $f$.
If they are similar, the transitivity function $R$ is applied to see if they are transitively similar.
If so, the clusters can be combined into a single cluster, with one representative
for all elements.
Otherwise, if the representatives are only similar but not transitive, members of either cluster \textit{might} be similar to the other representative.
To avoid missing these significant pairs, each cluster is compared against the other's representative and the similar elements are duplicated in the other cluster.
Finally, if the representatives are not similar, both clusters remain unchanged.
The result is a set of one or two clusters that are no longer mergeable.

\begin{algorithm}[!t]
  \caption{Cluster Set Merge}\label{algo:newmerge}
  \begin{algorithmic}[0]
    \Procedure{Merge}{$cs1, cs2$} \Comment{merge cluster set 1, 2}
      \State $newClusterSet \gets \emptyset$
      \For{$cluster1$ in $cs1$}
        \For{$cluster2$ in $cs2$}
          \State{\textbf{if} ($cluster2.HasBeenMerged$) $continue$}
          \State{$s \gets f(cluster1.rep, cluster2.rep)$}\Comment{Similarity}
          \State{\textbf{if} ($s < T$) $continue$}
          \If{$cluster2.IsTransitive(cluster1)$}
            \State{$cluster2.Objs.append(cluster1.Objs)$}
            \State{$cluster1.HasBeenMerged \gets True$}
            \State{$break$}
          \ElsIf{$cluster1.IsTransitive(cluster2)$}
            \State{$cluster1.Objs.append(cluster2.Objs)$}
            \State{$cluster2.HasBeenMerged \gets True$}
          \Else
            \State{$ExchangeSimilar(cluster1, cluster2)$}
          \EndIf
        \EndFor
        \If{$!cluster1.IsMerged$}
          \State{$newClusterSet.append(cluster1)$}
        \EndIf
      \EndFor
      \For{$cluster2$ in $cs2$} \Comment{add unmerged clusters}
        \If{$!cluster2.HasBeenMerged$}
          \State{$newClusterSet.append(cluster2)$}
        \EndIf
      \EndFor
      \State{\textbf{return} $newClusterSet$}
    \EndProcedure
  \end{algorithmic}
\end{algorithm}

Merging can also be applied to two \textit{sets} of clusters.
Algorithm~\ref{algo:newmerge} describes the process in detail.
Each cluster in the first set (\textit{cs1}) is compared to and possibly merged with every cluster in the second set~(\textit{cs2}).
For each cluster pair, the process described above is applied.
Finally, all unmergeable clusters are returned in a new set.

\subsection{ClusterMerge Algorithm}

\begin{figure}[t]
    \centering
    \includegraphics[width=0.8\columnwidth]{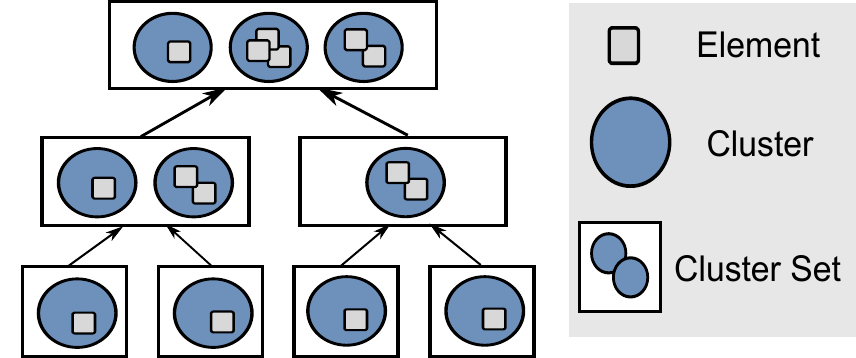}
    \caption{ClusterMerge algorithm. Elements are placed in trivial clusters which are then merged until an unmergeable set remains.}
    \label{fig:mergetree}
\end{figure}

The ClusterMerge algorithm uses cluster merging to perform precise clustering.
Each element is initially placed in its own cluster as its representative and each cluster is placed in its own set.
Algorithm~\ref{algo:newmerge} is then applied to merge cluster sets in a bottom-up fashion as depicted in Figure~\ref{fig:mergetree}.

Algorithm~\ref{algo:bottomup} describes this bottom-up merge process.
To start, a new cluster set is created for each element, with a single cluster containing only that element.
These cluster sets are added to a FIFO queue of sets to merge (the \emph{setsToMerge} queue).
The algorithm pops two sets off the queue, merges them using Algorithm~\ref{algo:newmerge}, 
and pushes the resulting cluster set onto the queue.
The process terminates when only one set is left.
This algorithm forms the basis of our implementations further described in \S\ref{sec:impl}.

\begin{algorithm}[!t]
  \caption{ClusterMerge}\label{algo:bottomup}
  \begin{algorithmic}[0]
    \Procedure{BottomUpMerge}{$elements$}
    \State $setsToMerge \gets Queue()$
    \For{$e$ in $elements$} 
      \State $setsToMerge.push(new~ClusterSet(e))$
    \EndFor
    \While{$setsToMerge.size() > 1$}
      \State $cs1 \gets setsToMerge.pop()$
      \State $cs2 \gets setsToMerge.pop()$
      \State $csNew \gets Merge(cs1, cs2)$ \Comment{merge sets cs1 \& cs2}
      \State $setsToMerge.push(csNew)$
    \EndWhile
    \State $finalSet \gets setsToMerge.pop()$ \Comment{final set of clusters}
    \EndProcedure
  \end{algorithmic}
\end{algorithm}

\subsection{Discussion}

With a complete\footnote{A \emph{complete} transitivity function correctly captures all transitive similarity in the data.} transitivity function, 
ClusterMerge will not miss any similar element pairs because all elements are implicitly compared against each other, either directly or implicitly via a transitive representative.
The chosen element remains representative of its cluster until it is (possibly) fully merged with another cluster. 
After that, transitivity ensures that subsequent similar elements will then also be similar to the new representative.
Therefore, even though cluster members are not necessarily \textit{transitively} represented by the cluster representative, the algorithm also ensures that those non-transitively similar elements retain their own cluster.

In reality, a complete and computationally efficient transitivity function rarely exists for non-trivial elements, so approximation is necessary, as for our motivating example of protein sequence clustering.
Incompleteness in the transitivity function can lead ClusterMerge to miss some similar pairs.
However, as is demonstrated in \S\ref{sec:eval}, even an approximate transitivity function can produce very good results.
This is also why transitivity is tested both ways in Algorithm~\ref{algo:newmerge}, since approximate transitivity is not necessarily symmetric. 

The threshold value $T$ is a parameter that would be chosen by an end user or domain expert to specify the desired degree of similarity between elements.
Users do not currently have influence over which elements are used as representatives, which are selected by the algorithm.

\subsection{Complexity}\label{sec:complexity}
The worst-case complexity of ClusterMerge is $O(n^2)$, however this is a fairly strict upper bound.
Consider the tree structure formed by the cluster set merges, which has a depth of $log_2 n$, where $n$ is the number of elements to be clustered.
At the first layer, there are $n/2$ merges possible, each comparing two clusters of one element each.
At there second layer, there are $n/4$ merges, each comparing a worst-case total of 4 clusters (if no full clusters were merged in the layer above).
Generalizing this pattern we obtain

\begin{equation*}
n/2\times1^2 + n/4\times2^2 + n/8\times4^2 \dots
\end{equation*}

which we can reduce to

\begin{equation*}
2n\sum\limits_{i=0}^{log_2 n} 2^i = 2n \cdot 2^{log_2 n + 1}-1 = 2n \cdot (2n) - 1 \approx n^2
\end{equation*}

However, when clusters are fully merged, there is a reduction in work at each level, leading to 
sub-$n^2$ performance.
In a more optimal case, assuming that at each step the merger of two cluster sets cuts the total number of clusters in half, 
complexity falls to $O(n \mathrm{log}n)$.
Actual complexity therefore depends on the amount of transitivity in the data being clustered.

\section{Protein Sequence Clustering}\label{sec:protclust}

Our motivation for this work is the problem of precise clustering for protein sequences.
Given their importance in biology, many specific algorithms for clustering sequences have been developed (\S\ref{sec:related}).
Fast algorithms however trade precision for speed and are not able to find a sufficient fraction of the similar sequence pairs in a dataset.

``Similar'' sequences in this domain indicate \textit{homologous} sequences/proteins/genes, 
with homology denoting the existence of a common ancestry between the sequences~\cite{patterson_homology_1988}. 
Homology within and across genomes can thus be used to propagate protein function annotations~\cite{gaudet_phylogenetic-based_2011, altenhoff_resolving_2012}.
In addition, homologous sequences can aid in the construction of phylogenetic species trees and the study of gene evolution~\cite{altenhoff_inferring_2012}.
As explained in \S\ref{sec:intro}, databases of homologous proteins are typically constructed using an expensive, all-against-all computation.
ClusterMerge can find a set of homologous pairs of proteins that closely approximates the set found by a full all-against-all approach, but at a much lower computational cost.

To apply ClusterMerge, we require a similarity function $f$, a clustering threshold $T$, and a transitivity function $R$.
The most accurate similarity function for proteins is a dynamic programming string matching algorithm, typically Smith-Waterman (S-W)~\cite{smith_identification_1981}, which is quadratic in the sequence length.
The \textit{score} produced by S-W is a weighted sum of matches, mismatches, and gaps in the optimal alignment, with the weights determined by a substitution matrix.
Our clustering threshold will be the same as in Wittwer et al.~\cite{wittwer_speeding_2014}, a S-W score of 181 using the PAM250 substitution matrix~\cite{dayhoff_chapter_1978}.
The transitivity function is constructed in a similar way to the incremental greedy clustering~\cite{wittwer_speeding_2014} and merits some additional explanation.

\begin{figure}[]
    \centering
    \includegraphics[width=\columnwidth]{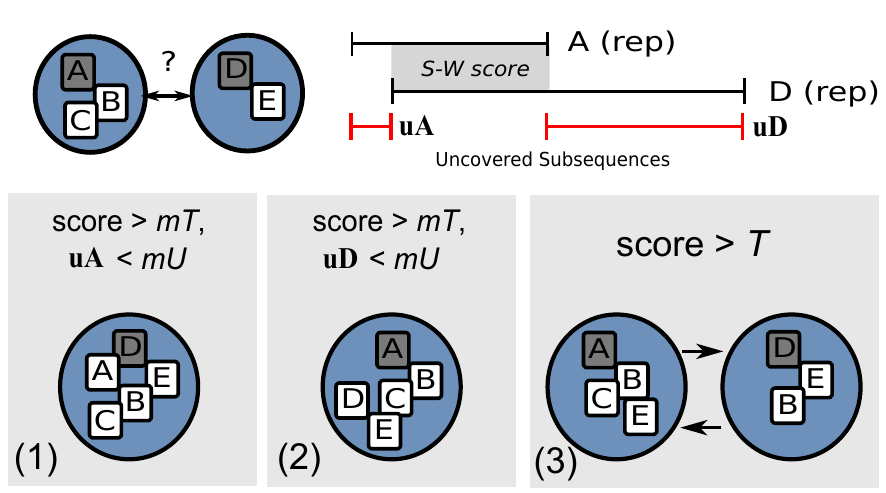}
    \caption{Transitivity function illustration for protein sequence clustering with ClusterMerge. }
    \label{fig:subseq_homology}
\end{figure}

Protein sequence alignment does have a transitive property, however S-W is a \textit{local} alignment algorithm, meaning that it may not include or ``cover'' all residues (individual amino acids) in both sequences, especially when the sequences are of different lengths.
If a sequence is clustered with a representative that does not completely cover it when aligned, the uncovered subsequence will go unrepresented.
This may cause \textit{subsequence homologies} to be missed.

Therefore, subsequence homologies must be taken into account when designing a transitivity function for proteins.
Figure~\ref{fig:subseq_homology} illustrates the transitivity function we use, through the example of merging two clusters with representative sequences A and D.
Depending on the size of each sequence and the alignment, there may be a number of uncovered residues in each sequence, shown as \textbf{uA} and \textbf{uD} in Figure~\ref{fig:subseq_homology}.
To fully merge the clusters, the alignment score between A and D must be greater than $mT$, the full merge threshold, a parameter.
In addition, the number of uncovered residues in one of the sequences must be less than parameter $mU$ 
(\textit{maximum uncovered}), to ensure that homologous subsequences are not missed.
For example, representative D has a large uncovered subsequence in Figure~\ref{fig:subseq_homology}, which representative A would not be able to transitively represent. 
Transitivity does not apply in this case.
However, D nearly completely covers A, and assuming \textbf{uA} is less than $mU$, transitivity would apply and D could transitively represent A.
The cluster of A would then be fully merged into the cluster of D, with D representing all sequences (situation (1) in Figure~\ref{fig:subseq_homology}).

\section{Parallel ClusterMerge}\label{sec:impl}

There are several opportunities for parallelism inherent in ClusterMerge, which we will use to construct efficient systems for both shared-memory and distributed environments.
Since the designs for shared-memory and distributed systems differ slightly, we will refer to the shared-memory design as Shared-CM and the distributed design as Dist-CM. 

The obvious parallelism in ClusterMerge is that smaller sets near the bottom of the tree can be merged in parallel.
In general, as long as there are sets of clusters to be merged in the setsToMerge queue, threads can pop two sets, merge them, and push the result back onto the queue.
These operations are independent and can be processed in parallel.

However, after many merges, only a few large sets remain. 
The ``tree-level'' parallelism is no longer sufficient to keep system resources occupied, and, in fact, the final set merge is always sequential.
Therefore, merges of individual sets must be parallelized, which is also necessary because the sets can grow to be very large.

\begin{figure}[]
    \centering
    \includegraphics[width=\columnwidth]{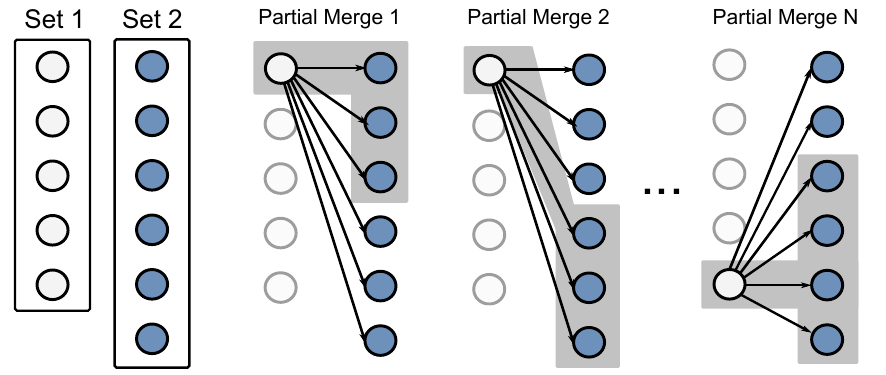}
    \caption{A merge of two large cluster sets is split into partial merges. Threads (or remote workers) can then
    simultaneously process a merge of two sets.}
    \label{fig:split_partial}
\end{figure}

Shared-CM and Dist-CM both use the same technique to split large set merges into smaller work items called \textit{partial merges}.
Consider merging two cluster sets, \textit{Set 1} and \textit{Set 2} (Figure~\ref{fig:split_partial}).
A \textit{partial merge} merges a single cluster from \textit{Set 1} into a subset of the clusters of \textit{Set 2}.
Threads or remote workers can execute these partial merges in parallel by running the full inner loop of Algorithm~\ref{algo:newmerge}.
This allows the system to maintain a consistent work granularity by scheduling a similar number of element comparisons in each partial merge.
Load is then evenly balanced, preventing stragglers and leading to better efficiency.
Shared-CM and Dist-CM differ only in how they coordinate synchronization of the results of partial merges.

\subsection{Shared-Memory}

Shared-CM is designed to be run on a typical commodity multicore computer.
Shared-CM splits set merges into partial merges as described above and allows threads to update the clusters in each set in place.

Consider a thread executing a partial merge, where a cluster from \textit{Set 1} is being merged into some clusters in \textit{Set 2}.
While our thread has exclusive access to the cluster from \textit{Set 1}, it has no such guarantee for the clusters in \textit{Set 2}.
Concurrent modifications, including removal of clusters and creation of new ones, can happen because of partial merges of other items from \textit{Set 1}.

Shared-CM uses locking to prevent races.
The merging logic is the same as in Algorithm~\ref{algo:newmerge}, however clusters of the second set are locked before being modified in-place.
The final merged set is simply the remaining clusters of \textit{Set 1} and \textit{Set 2} that have not been fully merged.
Ordering is not guaranteed and the process sacrifices determinism, but the significant pair recall is the same as a deterministic execution.


\begin{figure}[t]
    \centering
    \includegraphics[width=\columnwidth]{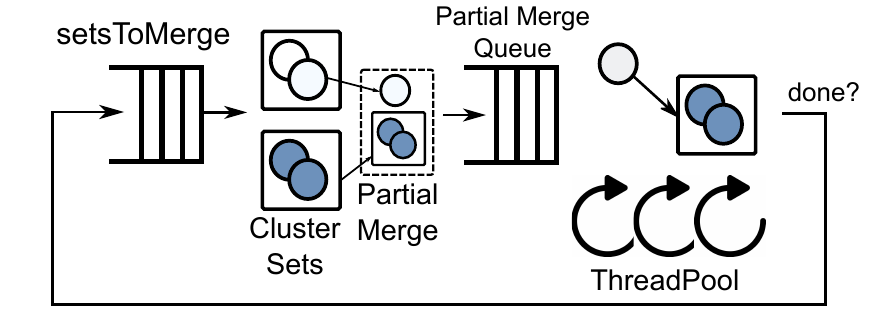}
    \caption{High-level architecture of Shared-CM.}
    \label{fig:sharedcm_arch}
\end{figure}

Figure~\ref{fig:sharedcm_arch} illustrates the system design.
A coordinating thread pops two sets off the \emph{setsToMerge} queue.
The merge is divided into partial merges as described above, which are inserted into a partial merge queue.
A pool of worker threads then process the partial merges.
Once the partial merges for a set merge are completed, the coordinating thread collects remaining clusters from both sets into a merged set and pushes it onto the queue.

As long as there are sets remaining to be merged, partial merges can be scheduled and all processors on the machine kept busy.
Multiple cluster set merges can also be split into partial merges and executed simultaneously.
Shared-CM can scale nearly linearly across cores, with full experimental results detailed in \S~\ref{sec:evalsm}.

\subsection{Distributed}

While locking works well in a multicore computer, it would limit scalability on a distributed cluster.
Instead, Dist-CM ensures that any processing sent to remote workers is fully independent.
Workers therefore have no communication with each other and only communicate with a central controller to get more work, resulting in a very scalable system.
Dist-CM is a controller-agent distributed system.
The \textit{controller} is responsible for managing the shared state of the computation, while the majority of the computing is performed by \textit{remote workers}.

Dist-CM uses several techniques to control the size of an average work item to prevent load imbalance and enable efficient scaling.
First, batching is used to group small cluster sets together as a single work item.
This provides each remote worker with a unit of computation that will not be dwarfed by communication overhead.
Batches are executed by a remote worker, and the resulting cluster set is returned to the controller and pushed back into the \emph{setsToMerge} queue.
Batching is important for the early phase of computation, where each set is small and requires little computation.

For larger merges near the top of the tree, Dist-CM uses partial merges in much the same manner as Shared-CM to maintain a consistent work item granularity.
Because there is no inter-worker communication, the controller is responsible for managing partial merge results as they are returned.
Recall that each partial merge work item merges a single cluster from \textit{Set 1} into a subset of clusters of \textit{Set 2}.
The result of a partial merge executed by a remote worker is then a set containing some clusters of \textit{Set 2}, with the single cluster from \textit{Set 1} possibly fully merged with one of them and/or some elements exchanged with some clusters.
If the single cluster was not fully merged, it will be included in the returned set.

For each outstanding merger of two cluster sets, the controller maintains a partially merged state of the final result, identified by an ID associated with all partial merges involved in its computation.
This partially merged state begins as simply both sets of clusters.
When a partial merge result is returned to the controller, it uses the ID to look up the associated partially merged state.
The controller will then update the partially merged state with the results of the returned partial merge, adding any elements to existing clusters and marking any fully merged clusters.
After processing the final partial merge for a given set merge, the merge is complete and the resulting set is constructed by simply combining non-fully merged clusters from both sets.

\begin{figure}[t]
    \centering
    \includegraphics[width=\columnwidth]{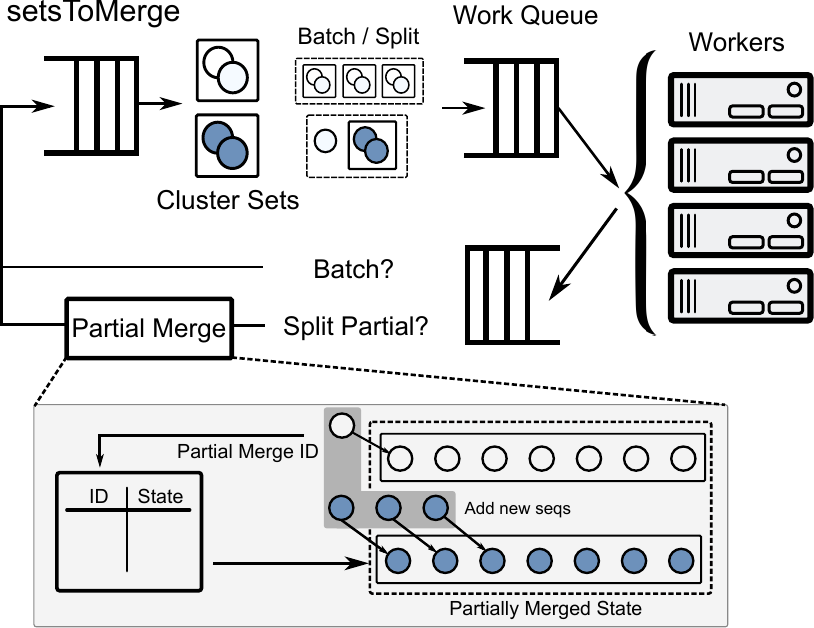}
    \caption{High-level architecture of Dist-CM.}
    \label{fig:distcm_arch}
\end{figure}

Figure~\ref{fig:distcm_arch} summarizes the design of Dist-CM.
Once again the \textit{setsToMerge} queue is loaded with single element cluster sets, but at the central controller.
A coordinating thread on the controller will pop two sets off the queue to merge together.
If the sets (in terms of total clusters) are smaller than a batch size parameter, the thread will pop more sets until it has sets whose total number of clusters is equal to or greater than the batch size parameter.
These sets are compiled into a batch work item and pushed into a central work queue.
If the sets popped by the coordinating thread are large, they are split into partial merges, with the number of sequences in each cluster taken into account to evenly size each request.
This dynamic load balancing keeps straggling in remote workers to a minimum, and is important in achieving good scaling.
Partial merges are then are pushed into the central work queue as individual work items.
The central work queue feeds a set of remote worker nodes.

Results from workers are returned to the controller and either pushed back to the \textit{setsToMerge} queue if a batch result (which is a complete cluster set), or used to update an associated partially merged state if it was a partial merge result.
If the partially merged state was completed by the work item in question, the now-complete set is pushed to the \textit{setsToMerge} queue.
The process is complete when the final set of the merge tree is complete.

The trade-off inherent in this design is that Dist-CM does more work than necessary in exchange for zero communication among workers.
A cluster in a partial merge will continue to be merged into clusters in the set by Dist-CM even if they were fully merged away in other workers.
As a result, Dist-CM can perform slightly more work than Shared-CM and can occasionally add the same elements to the same cluster (these duplicates are easily removed by the controller).
This trade-off leads to Dist-CM being about 17\% slower than Shared-CM when using a single remote worker in our application of protein sequence clustering.

Scalability can be adversely affected by average latency or amount of work in a single work item.
Very small work items will have communication overheads that may dwarf the actual computation.
Very large work items can cause stragglers and load imbalance that can leave processors idle.
Early versions of Dist-CM operated without dynamic sizing of partial merges; each one merged a single cluster into the entirety of the other set.
This led to massive load imbalance and long idle periods as large merges were completed.
Dynamic sizing of partial merges was crucial to ensure proper load balance and minimize stragglers, improving scaling efficiency by almost $3\times$.

Furthermore, unbalanced work distribution can cause stragglers as well, if some workers locally queue more work than others. 
To avoid this, we switched from a round-robin work distribution method to a Join-Idle-Queue~\cite{lu_join-idle-queue:_2011} approach in which workers inform the controller when they need more work.
This keeps all workers busy so long as work is available, while limiting worker-local queuing.

\subsection{Optimizations}

Several important optimizations enable efficient scaling of Dist-CM.
In early versions, the controller sent whole sets with each partial merge, which nearly saturated available network bandwidth.
Communication overhead was greatly reduced through several techniques.
First, each worker replicates the sequence dataset and refers to sequences by a 4 byte index.
Actual sequence data is never transferred, and even large clusters with thousands of sequences only require a few kilobytes.
Second, one of the sets in a series of partial merges is cached on each worker, so it is only transferred over the network once.
Finally, the results of partial merge are returned as diffs, i.e., only newly added sequences in each cluster.

\section{Evaluation}\label{sec:eval}

We evaluate several aspects of ClusterMerge and its implementations applied to protein sequence clustering.
Both Shared-CM and Dist-CM variants are evaluated in this section.
Both implementations are written in C++ and compiled with GCC 5.4.0.
To compute sequence similarities, we use the Smith-Waterman library SWPS3~\cite{szalkowski_swps3_2008}.

We use two datasets for our evaluation.
One is a dataset of 13 bacterial genomes extracted from the OMA database~\cite{altenhoff_oma_2018}, 
a total of 59013 protein sequences (59K dataset). 
This is the same dataset used by Wittwer et al., which allows comparison with their implementation.
The second dataset is a large set of eight genomes from the QfO benchmark totaling 90557 sequences (90K dataset).
Although these are a small fraction of the available databases, each represents billions of possible similar pairs, taking many hours to evaluate in a brute-force manner.

Our tests are performed using servers containing two Intel Xeon E5-2680v3 running at 2.5 GHz 
(12 physical cores in two sockets, 48 hyperthreads total), 256 GB of RAM, running Ubuntu Linux 16.04.
The distributed compute cluster consists of 32 servers (768 cores), a subset of a larger, shared deployment. 
These are connected via 10 Gb uplinks to a 40GbE-based IP fabric with 8 top-of-rack switches and 3 spine switches.
The dataset is small enough such that a local copy can be stored on each server.
In fact, even large protein datasets are easily stored on modern servers. 
For example, the complete OMA database of 14 million protein sequences fits within 10GB, a fraction of modern server memory capacity.

Our baseline for clustering comparisons is the incremental greedy precise clustering of~\cite{wittwer_speeding_2014}, which is the only clustering method that can achieve an equivalent level of similar pair recall.

\subsection{Clustering and Similar Pair Recall}\label{sec:evalclustering}

For consistency, our clustering threshold is the same as the incremental greedy precise clustering in Wittwer et al.~\cite{wittwer_speeding_2014}, a Smith-Waterman score of 181.
The threshold is low, but this is necessary to find distant homologs.
After ClusterMerge identifies clusters, an intra-cluster, all-against-all comparison is performed, in which the sequence pairs within a cluster are aligned using Smith-Waterman.
Those with a score higher than the clustering threshold are recorded as a similar pair.
For our datasets, the number of actual similar pairs is small compared to the number of potential similar pairs (e.g. 1.2 million actual versus 1.74 billion potential), leading to relatively few alignments to complete this stage.
Biologists may perform additional alignments to derive an optimal alignment with respect to different scoring matrices, however this is orthogonal to the concerns of this paper.

\begin{figure}[t]
    \centering
    \includegraphics[width=\columnwidth]{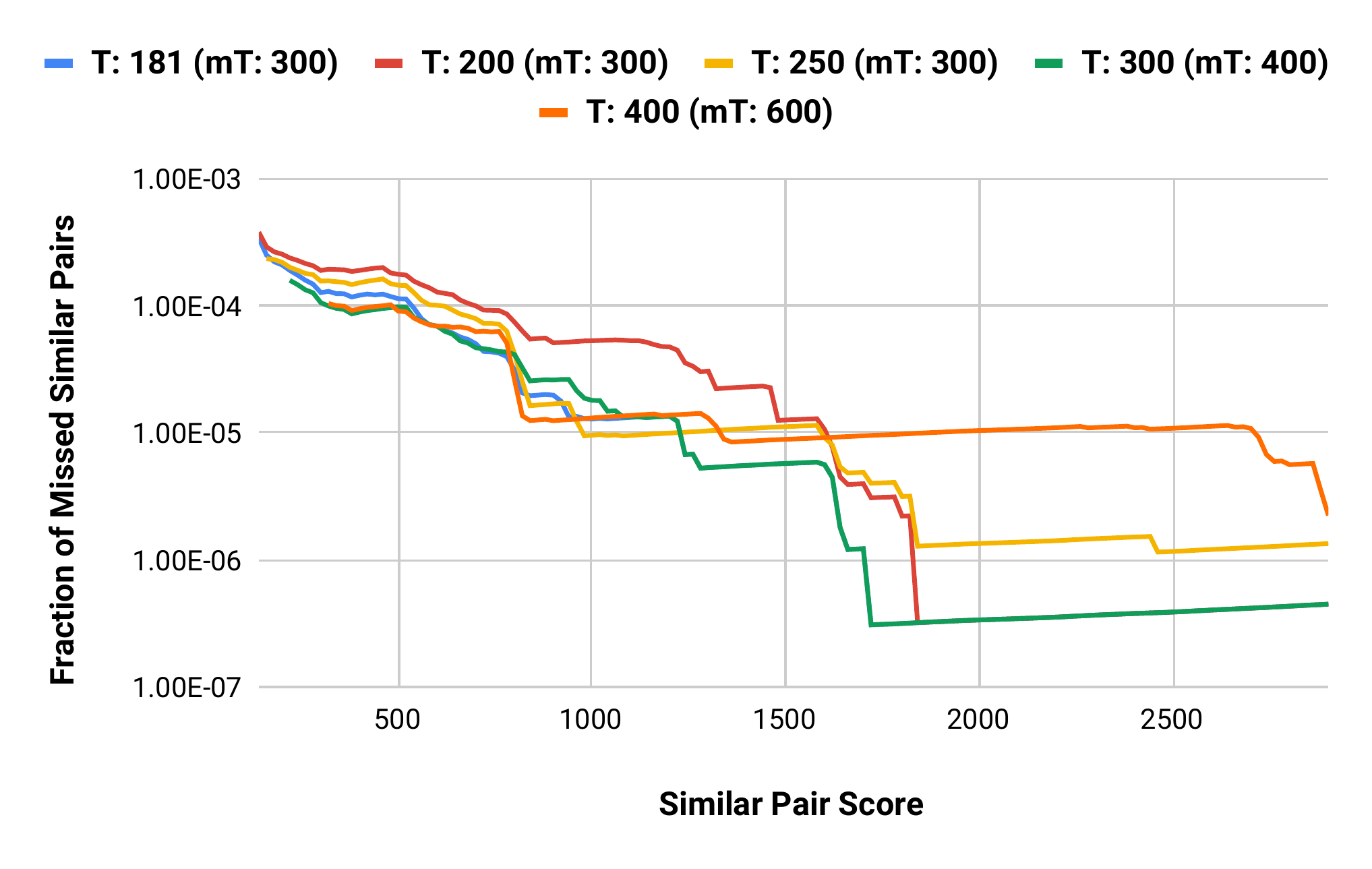}
    \caption{Cumulative fraction of missed pairs reaching at least a certain similarity score, as the clustering threshold $T$ and fully merge threshold 
    $mT$ are varied ($mU = 15$). ClusterMerge shows a low sensitivity to small parameter variations, while most missed similar pairs 
    remain low-scoring ones.}
    \label{fig:missed_pair_sweep}
\end{figure}

Recall is the percentage of ground truth pairs found by our systems.
Ideal recall is 100\%.
Both Shared-CM and Dist-CM ClusterMerge, using a minimum full merge score ($mT$) of 250 and a max uncovered residues ($mU$) of 15, produce clusters with a recall of $99.8\pm 0.01$\%.
Recall variability is negligible and is due to the non-determinism of parallel execution.
Of the pairs missed by ClusterMerge, very few were high scoring pairs.
The median score of a missed pair is 191 and the average score of a missed pair is 235.
These values are very close to the cluster threshold itself (in contrast to high scoring pairs, which can be greater than 1000), indicating that these are not likely biologically ``important'' pairs (Figure~\ref{fig:missed_pair_sweep}).
ClusterMerge misses only a handful of high scoring pairs, around one millionth of total significant pairs, as seen in Figure~\ref{fig:missed_pair_sweep}.

In clustering the 59K sequence dataset, ClusterMerge performs approximately 871 million comparisons.
By contrast, the full, all-against-all comparison requires approximately 1.74 billion comparisons, showing that ClusterMerge reduces comparisons by nearly 50\%.

In terms of the clusters themselves, ClusterMerge generates a similar clustering profile as incremental greedy clustering~\cite{wittwer_speeding_2014}, with a total of 33,562 clusters.
In each, the vast majority of clusters contain between 1 and 4 sequences, with a few large clusters (33\% of clusters contain more than 10 sequences, 8\% of clusters contain more than 100 sequences, 0.5\% of clusters contain more than 1000 sequences).
ClusterMerge generates slightly larger outliers, with its largest cluster containing approximately 1500 sequences as opposed to the greedy method's largest cluster of around 1150 sequences.

Figure~\ref{fig:missed_pair_sweep} shows that ClusterMerge and our transitivity function are relatively insensitive to parameter variations.
Lower clustering thresholds $T$ and lower full merge thresholds $mT$ generally lower the number of missed similar pairs, although the absolute percentage of missed pairs remains extremely low, with the majority being low-scoring pairs.

\subsection{Multicore Shared-Memory Performance}\label{sec:evalsm}
\begin{figure}[t]
    \centering
    \includegraphics[width=0.9\columnwidth]{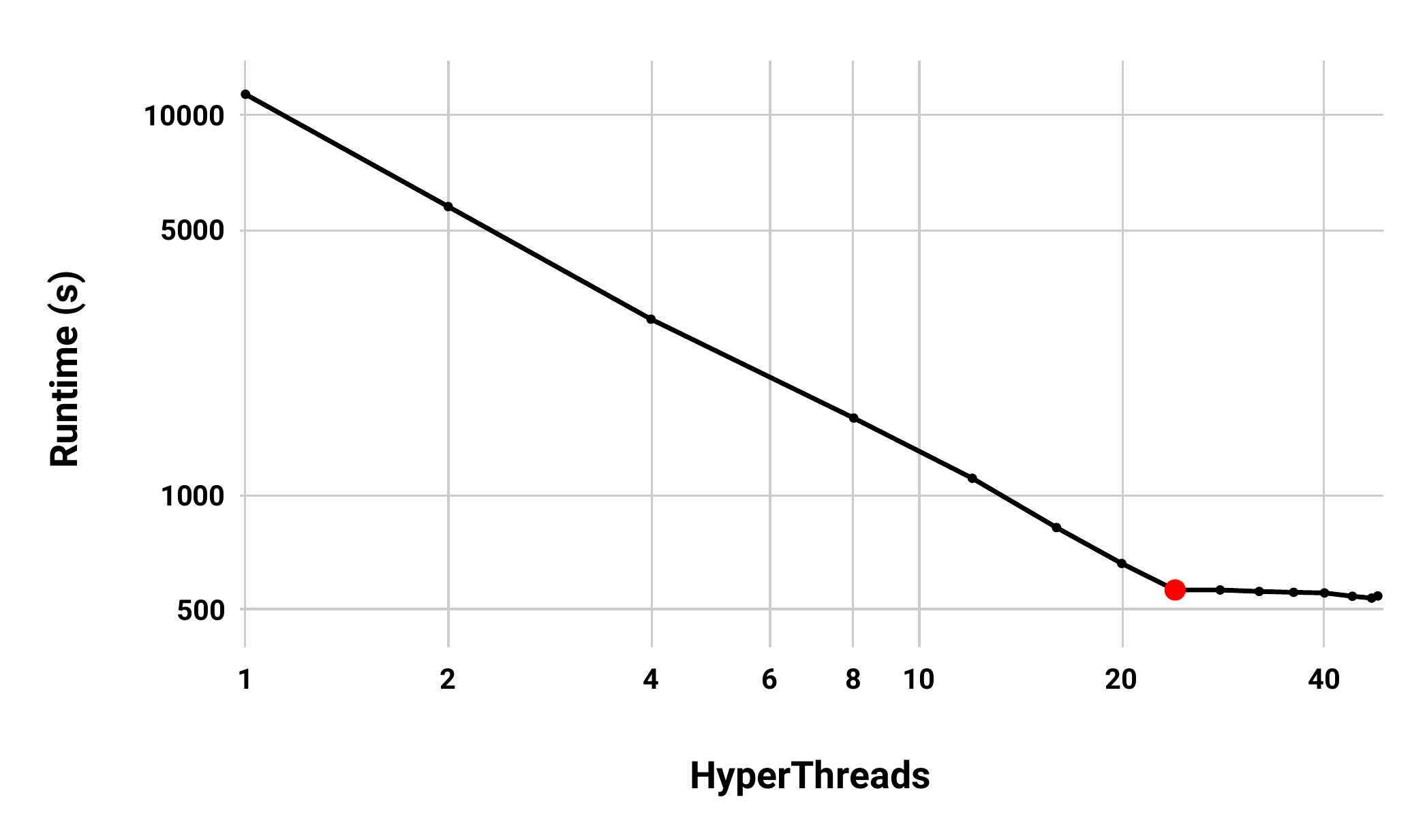}
    \caption{Scaling of Shared-CM up to 48 threads. Scaling is nearly linear up to all 24 physical cores, while hyperthreading provides no benefit.}
    \label{fig:thread_scaling}
\end{figure}

In this section, we evaluate how well Shared-CM performs on a single multicore node.
This experiment uses a reduced dataset of 28600 sequences, to reduce runtimes at low thread counts.
Figure~\ref{fig:thread_scaling} shows the total runtime decreases as we increase the number of threads.
Shared-CM achieves near linear scaling --- profiling with Intel VTune indicates little or no lock contention.
Memory access latency and NUMA have no effect as the workload is compute bound.

Note, however, that scaling is linear only on physical cores.
The primary compute bottleneck is the process of aligning representative sequences using Smith-Waterman,
which processes data that fits in the L1 cache and is able to saturate functional units with a single thread.
Therefore, hyperthreading provides no benefit.

The only major impediment to perfect scaling is some loss of parallelism before the last and second-last merges, since the second-last merge must be fully completed before work for the last merge can start to be scheduled.

Shared-CM with a single thread clusters the bacteria dataset in 31905 seconds, compared to 1486 seconds with 24 threads, a speedup of \sharedSpeedup.
To compare with incremental greedy clustering, we run Wittwer's single-threaded code~\cite{wittwer_speeding_2014} on our machine with the same dataset, resulting in a runtime of \wittwerBaseline seconds.
Shared-CM is approximately 2.8$\times$ faster on a single core, and \sharedSpeedupOverGreedy faster using all cores.

\subsection{Distributed Performance}\label{sec:evaldist}

\begin{figure}[t!]
	\centering
	\begin{subfigure}[b]{\columnwidth}
		\includegraphics[width=\columnwidth]{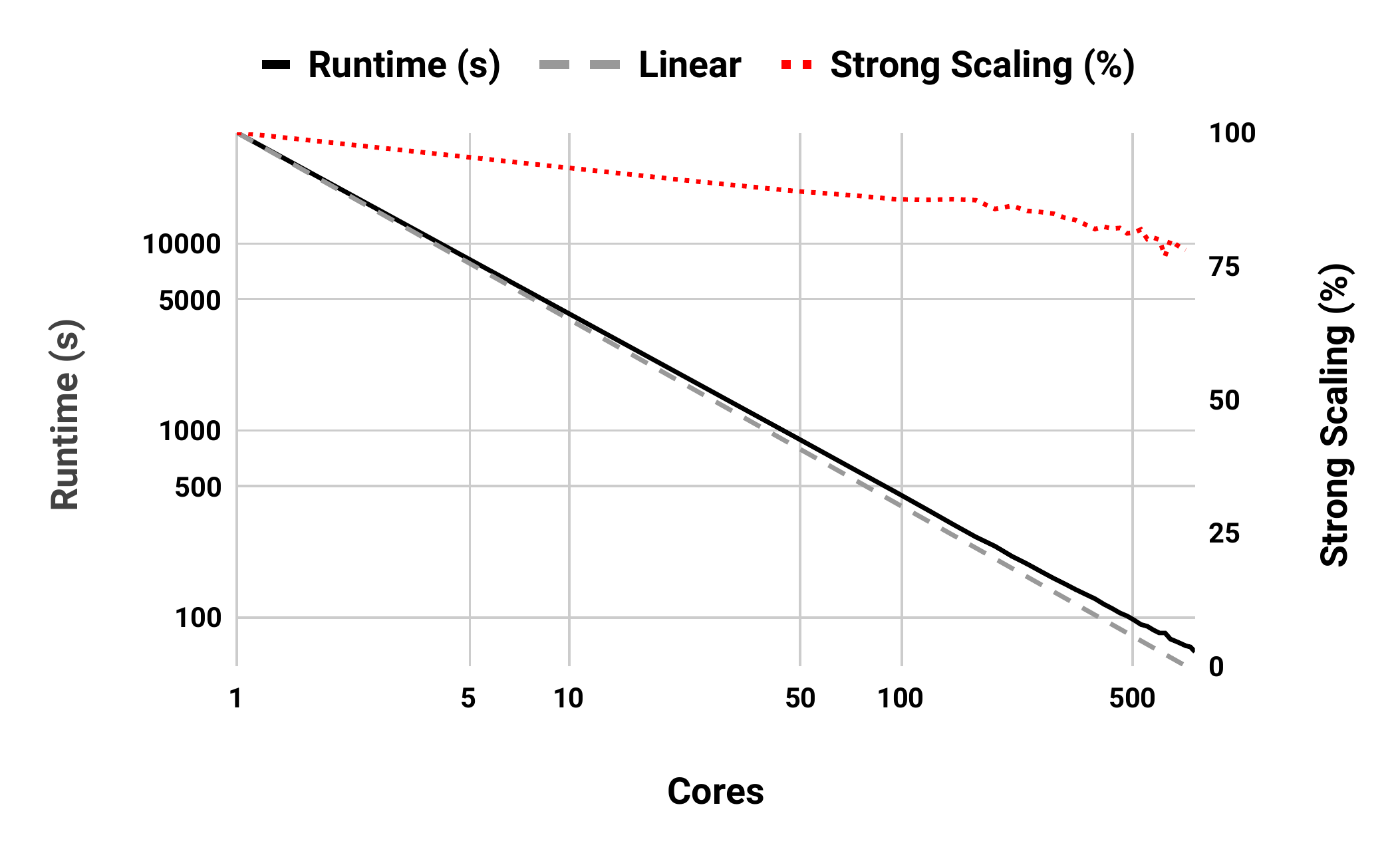}
		\caption{59K dataset, 79\% scaling efficiency.}
        \label{fig:node_scaling_small}
	\end{subfigure}
	\begin{subfigure}[b]{\columnwidth}
		\includegraphics[width=\columnwidth]{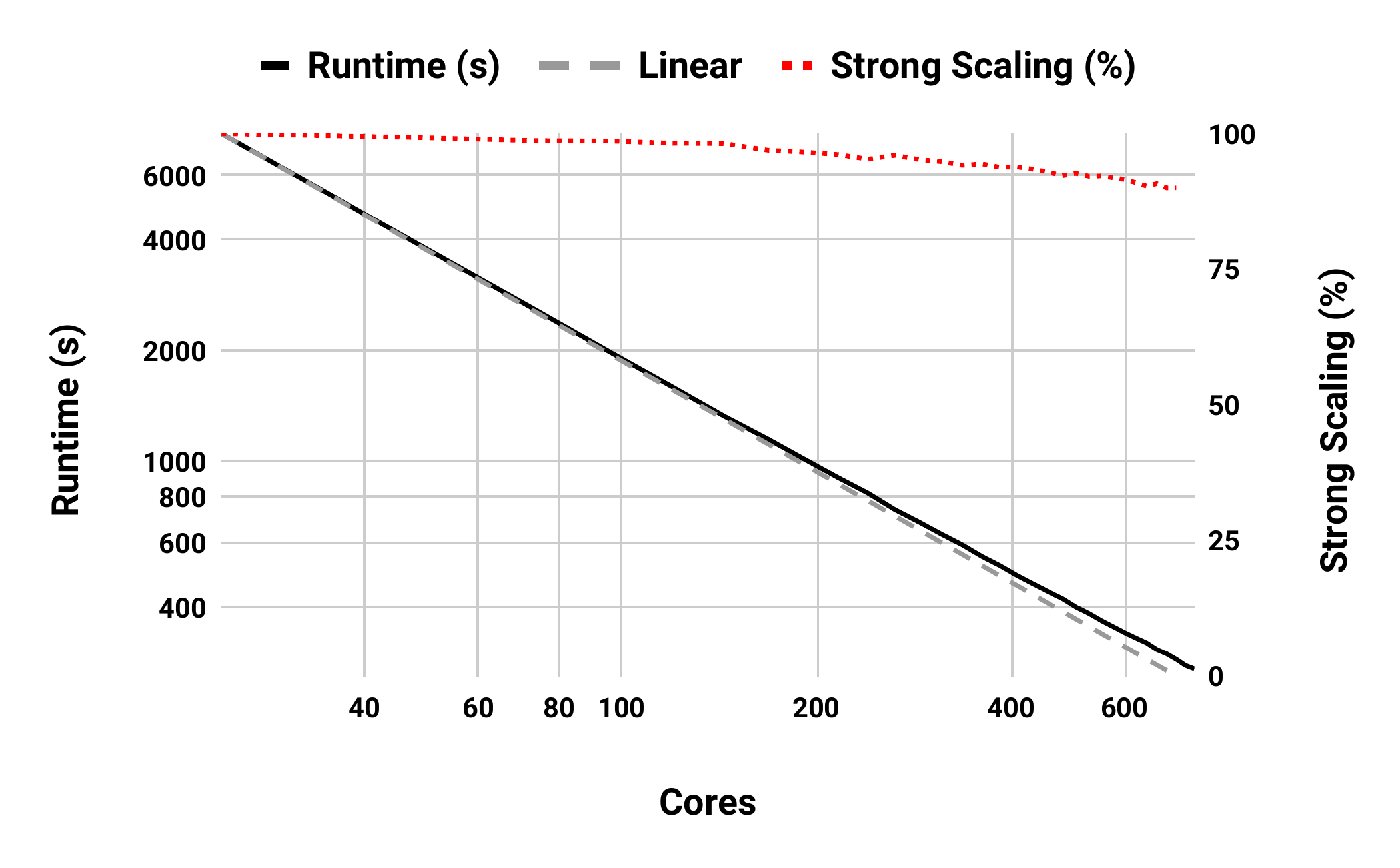}
		\caption{90K dataset, 90\% scaling efficiency.}
        \label{fig:node_scaling}
	\end{subfigure}
	\caption{Scaling of Dist-CM over 32 servers (768 cores).}
	\label{fig:dist_scaling}
\end{figure}

Dist-CM allows us to scale ClusterMerge beyond a single server.
To evaluate the scaling of Dist-CM, we hold the dataset size constant and vary the number of servers 
used to process work items (batches or partial merges), otherwise known as \textit{strong scaling}.
The baseline single core runtime for Dist-CM clustering the 59K dataset is \distBaseline seconds. 
Figure~\ref{fig:node_scaling_small} shows that on 32 nodes (768 cores) Dist-CM clusters the dataset in 65 seconds, resulting in a speedup of \distSpeedup.
Strong scaling efficiency at 768 cores is \distSmallStrongScaling.
Compared to single-threaded incremental greedy clustering~\cite{wittwer_speeding_2014}, Dist-CM is 2.27$\times$ faster 
using a single core, and \distSpeedupOverGreedy faster using the full compute cluster.

The reason for sublinear scaling is essentially the same as with Shared-CM --- around the last few merges 
of cluster sets, work scheduling may halt as the system waits for an earlier merge to finish before being able to schedule more work.
There will always be some small portion of sequential execution, so perfect scaling is impossible by Amdahl's Law.

\begin{figure}[t]
    \centering
    \includegraphics[width=0.95\columnwidth]{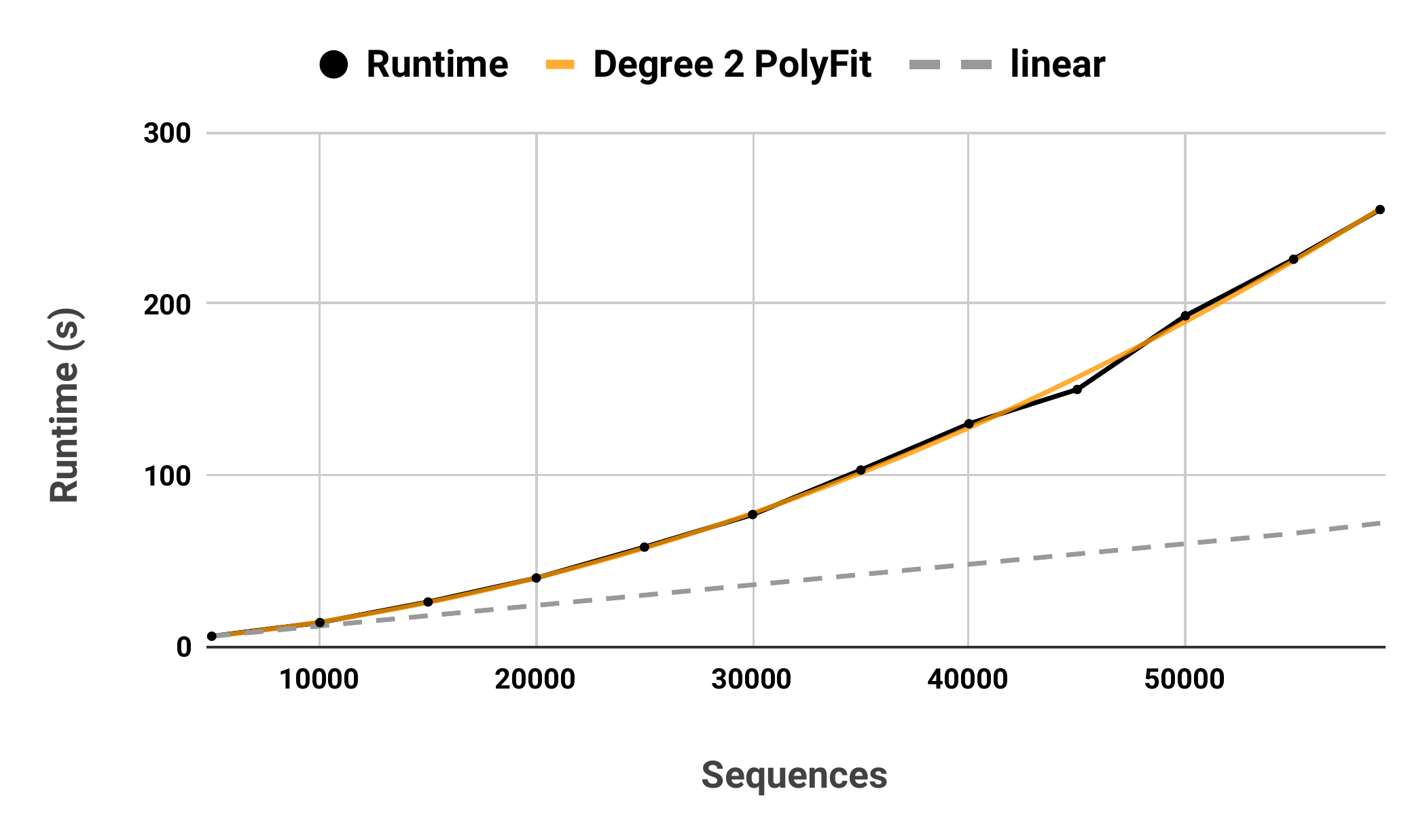}
    \caption{Workload scaling of Dist-CM. }
    \label{fig:workload_scaling}
\end{figure}
\begin{figure}[t]
    \centering
    \includegraphics[width=\columnwidth]{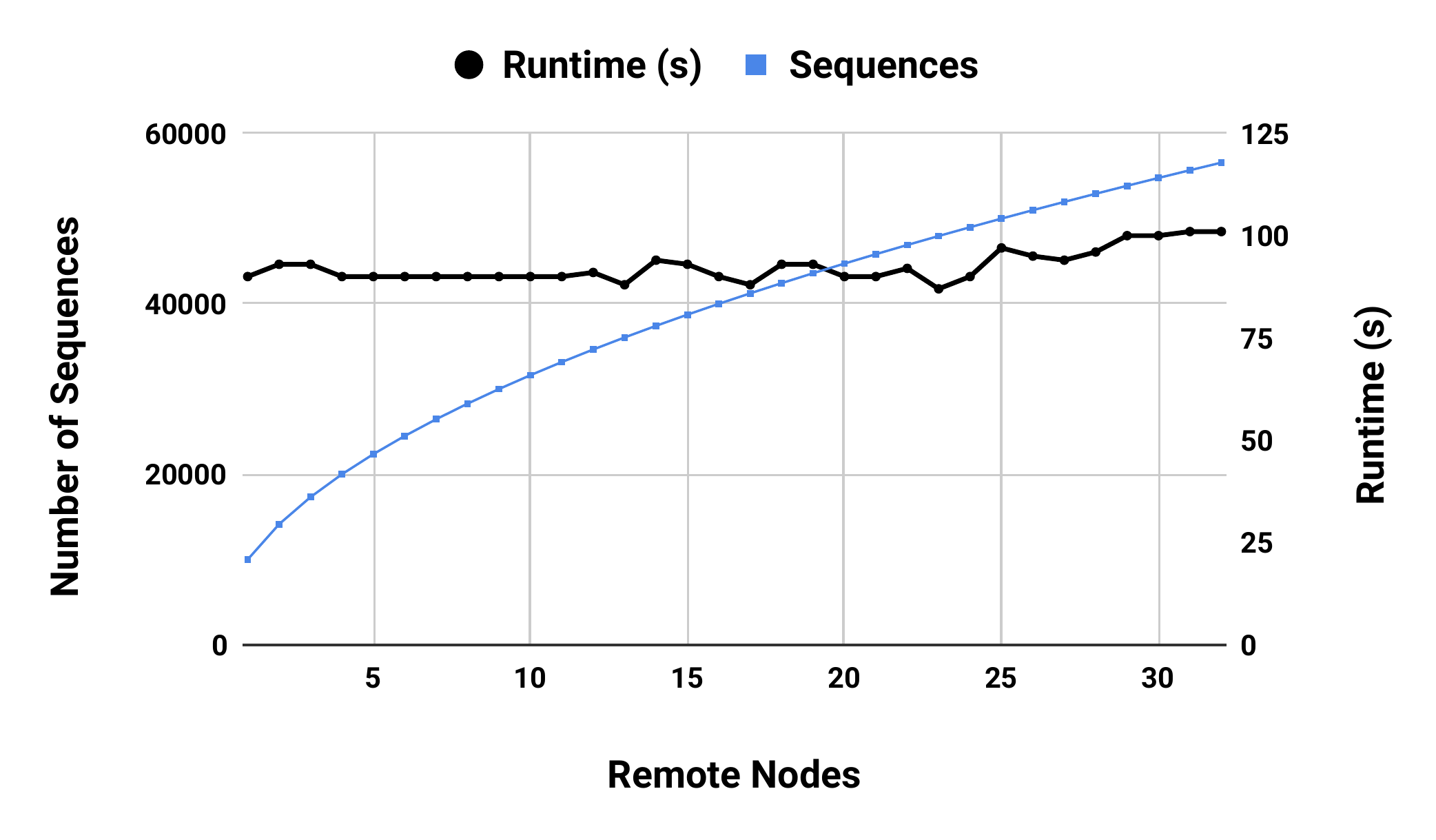}
    \caption{Dist-CM weak scaling over a 32 node (768 core) cluster. Nearly 100\% efficiency at 32 nodes.}
    \label{fig:node_scaling_weak}
\end{figure}

That being said, this sequential section is proportionally lower with larger datasets.
Figure~\ref{fig:node_scaling} shows Dist-CM strong scaling when clustering the larger 90K sequence dataset.
The scaling is much more efficient (90\% at 32 nodes), with a speedup of 28.7$\times$ relative to one server node.

In addition, we perform a weak scaling experiment in which we vary the amount of work in proportion to the number of nodes.
Because our dataset is evolutionarily diverse and has relatively low levels of transitivity, ClusterMerge is closer to $O(n^2)$ in the number of sequences.
The amount of actual work increases quadratically with the number of sequences.
Figure~\ref{fig:workload_scaling} clearly shows this by varying the number of sequences that Dist-CM clusters using 10 worker nodes. 
The runtime curve fits almost exactly to a degree two polynomial.
Therefore, for our weak scaling experiment, we vary the number of sequences at each step by a square root factor to maintain a proportional increase in workload.
Figure~\ref{fig:node_scaling_weak} shows the results, again while clustering using 1 to 32 nodes.
Runtime remains nearly constant throughout, indicating a weak scaling efficiency of 95-100\%.
We thus expect that Dist-CM will be able to cluster much larger datasets while maintaining high scaling efficiency.

\subsection{Effect of Dataset Composition}
As noted in section \S\ref{sec:complexity}, complexity and therefore runtime depend on how many clusters can be fully merged at each level of the tree.
If the transitivity function accurately represents similar elements, the number of full merges at each level is primarily affected by the number of transitively similar elements in a dataset. 
More transitively similar elements will result in more complete cluster merges, bringing runtime complexity closer to the $O(n\mathrm{log}n)$ optimum.

For protein clustering, the dataset with 13 bacterial genomes has a relatively low number of transitively similar sequences since the species are genetically very distant (more distant than human and plants).
Given a set of more closely related genomes, with  more transitively similar sequences, we would expect ClusterMerge to generate fewer clusters and run much faster.
To test this hypothesis, we clustered a third dataset of more closely related \textit{Streptococcus} bacteria genomes, consisting of 33 genomes (69648 sequences, similar to the other dataset).

Using Shared-CM with 48 threads, the clustering is completed in 283 seconds, producing 10500 clusters. 
As predicted, clustering is much faster than the 13 bacterial genome dataset (1486 seconds) as the number of clusters is much lower. 
In addition, ClusterMerge again produced a high recall of 99.7\% of similar pairs relative to a full all-against-all.
\section{Future Work}\label{sec:future}

Both of our implementations, Shared-CM and Dist-CM, perform and scale well.
However many improvements are possible.
Dataset size may expose limits to the current implementation.
Very large clusters may produce work items that are still too large, which may cause straggling.
Additional splitting beyond the current partial merge may be necessary.
Extreme imbalance in cluster sizes between two sets to be merged may also require more creative scheduling of partial merges to avoid large variation in work item size.



For our application to proteins, the current computational bottleneck is the Smith-Waterman alignment function.
Runtimes could be improved with a more efficient S-W implementations.
We are actively investigating protein alignment-friendly S-W hardware implementations.
Similarly, more approximate or less precise alignment methods could be used, though this may come at the cost of precision.

\section{Conclusion}\label{sec:conclusion}

ClusterMerge is a parallel and scalable algorithm for precise clustering of elements.
When applied to protein sequences, ClusterMerge produces clusters that encompass 99.8\% of significant pairs found by a full all-against-all comparison, while performing 50\% fewer similarity comparisons.
Our implementations achieve speedups of \sharedSpeedup on a 24-core shared-memory machine and \distSpeedup on a cluster of 32 nodes (768 cores).
The distributed implementation of ClusterMerge for protein clustering can produce clusters \distSpeedupOverGreedy faster than a single-threaded greedy incremental approach.
ClusterMerge is open source and available~\cite{noauthor_clustermerge_nodate}.

Our hope is that ClusterMerge will help to form a comprehensive ``map'' of protein sequences.
In theory, clustering could proceed to a point where any given new protein sequence would be 
represented completely by a subset of existing clusters.
No new clusters would need to be added, and any new protein could be classified in $O(n\mathrm{log}n)$ time only.
However, it is not yet clear how many different genomes would be required to form such a map.
%

%
\bibliographystyle{ACM-Reference-Format}
\bibliography{refs}

\end{document}